\documentclass{emulateapj}
\def \be{\begin{equation}}
\def \ee{\end{equation}}
\def \bdm{\begin{eqnarray}}
\def \edm{\end{eqnarray}}
\begin{document}
\title{Perpendicular Diffusion of Energetic Particles in Noisy Reduced Magnetohydrodynamic Turbulence}
\author{A. Shalchi \& M. Hussein}
\affil{Department of Physics and Astronomy, University of Manitoba, Winnipeg, Manitoba R3T 2N2, Canada}
\email{andreasm4@yahoo.com \\ m\_hussein@physics.umanitoba.ca}
\begin{abstract}
Recently a model for noisy reduced magnetohydrodynamic turbulence was proposed. The latter model was already
used to study the random walk of magnetic field lines. In the current article we use the same model to investigate
the diffusion of energetic particles across the mean magnetic field. To compute the perpendicular diffusion
coefficient two analytical theories are used, namely the Non-Linear Guiding Center (NLGC) theory and the
Unified Non-Linear Transport (UNLT) theory. It is shown that the two theories provide different results
for the perpendicular diffusion coefficient. We also perform test-particle simulations for the aforementioned
turbulence model. We show that only the UNLT theory describes perpendicular transport accurately confirming
that the latter theory is a powerful tool in diffusion theory.
\end{abstract}
\section{Introduction}
In the current paper we explore perpendicular diffusion of energetic particles such as cosmic rays due to
the interaction with turbulent magnetic fields. The perpendicular diffusion coefficient is one of the elements
entering the cosmic ray transport equation. In general, the diffusion of energetic particles is important
to understand different processes in space and astrophysics. Some examples which were discussed more recently in the
literature are:
\begin{itemize}
\item The acceleration of particles due to turbulence (see Lynn et al. 2014).
\item Shock acceleration at interplanetary shocks (see Li et al. 2012, Wang et al. 2012).
\item Solar modulation studies (see Alania et al. 2013, Engelbrecht \& Burger 2013, Manuel et al. 2014, Potgieter et al. 2014).
\item The motion of cosmic rays in our own and in external galaxies (see Buffie et al. 2013, Berkhuijsen et al. 2013).
\item Diffusive shock acceleration in supernova remnants (see Ferrand et al. 2014).
\end{itemize}
In the current article we explore perpendicular transport analytically and numerically for a specific turbulence
model.

In the solar system, for instance, energetic particles interact with the solar wind plasma and, therefore, they are
scattered. Spatial diffusion is mainly caused due turbulent magnetic fields $\delta \vec{B}$. In addition to such fields
we also find an ordered magnetic field $\vec{B}_0$ which breaks the symmetry of the considered physical system.
Therefore, we have to distinguish between diffusion of particles along and across the ordered magnetic field which
can also be called the mean magnetic field.

Especially diffusion across this field, also called perpendicular diffusion, is very difficult to describe analytically
(see Shalchi 2009 for a review). More than a decade ago some progress has been achieved due to the development of the
Non-Linear Guiding Center (NLGC) theory of Matthaeus et al. (2003) and more recently the Unified Non-Linear Transport
(UNLT) theory was presented in Shalchi (2010). The latter theory contains the NLGC theory, the field line transport theory
of Matthaeus et al. (1995), and the quasi-linear theory of Jokipii (1966) as special limits. Furthermore, the theory
automatically provides a subdiffusive result for magnetostatic slab turbulence in agreement with the theorem on reduced
dimensionality (see Jokipii et al. 1993 and Jones et al. 1998) and computer simulations (see, e.g., Qin et al. 2002).

In order to compute the perpendicular diffusion coefficient based on the aforementioned transport theories, one has
to employ a certain turbulence model. Previous models for which the perpendicular diffusion coefficient was calculated
are the slab/2D composite model (sometimes called two-component model) and the Goldreich-Sridhar model (see, e.g.,
Tautz \& Shalchi 2011 and Shalchi 2013a). In the current paper we employ another model which was proposed recently,
namely the model of {\it Noisy Reduced MagnetoHydroDynamic (NRMHD) turbulence} of Ruffolo \& Matthaeus (2013).

In the present paper we explore perpendicular diffusion in NRMHD turbulence analytically and numerically. By doing this
we try to achieve the following:
\begin{enumerate}
\item We show how the field line random walk limit with the correct field line diffusion coefficient can be obtained
from the UNLT theory.
\item The first time we obtain the perpendicular diffusion coefficient of energetic particles for NRMHD turbulence. 
\item We test the validity of NLGC and UNLT theories by comparing them with test-particle simulations in order to
check our understanding of perpendicular diffusion.
\end{enumerate}

The remainder of this paper is organized as follows.  In Section 2 we briefly present the NLGC theory as well as the UNLT
theory. A discussion of the NRMHD turbulence model is given in Section 3. In Section 4 we compute the perpendicular diffusion
coefficient analytically and in Section 5 we use the simulations to test our analytical findings. We end with a short
summary and some conclusions in Section 6.
\section{Analytical Theories for Perpendicular Diffusion}
The analytical description of perpendicular diffusion is difficult (see Shalchi 2009 for a review) since
a quasilinear approximation is only valid in exceptional cases. A promising theory was proposed by Matthaeus et al. (2003)
which is called the Non-Linear Guiding Center (NLGC) theory. The latter theory was compared with test-particle simulations
and solar wind observations and agreement was often found (see, e.g., Matthaeus et al. 2003, Bieber et al. 2004).
However, there are also problems with the theory such as the fact that the theory does not provide subdiffusive transport
for slab turbulence\footnote{We like to emphasize that the subdiffusive behavior is an aspect of pure magnetostatic
slab turbulence and for this specific model NLGC theory doesn't work. For a slab/2D composite model, however, diffusion
should be recovered. For two-dimensional turbulence, NLGC theory should be valid. It is not our intention to criticize the
slab/2D model or any other model of magnetic turbulence.} (see Shalchi 2009, Tautz \& Shalchi 2011). Therefore, different extensions of the NLGC theory
were proposed. One example is the Extended Non-Linear Guiding Center (ENLGC) proposed by Shalchi (2006). This theory
was explicitly developed to handle perpendicular transport in slab/2D composite turbulence and provides the
correct subdiffusive behavior for the pure slab case. Alternative approaches were proposed thereafter (see, e.g.,
Qin 2007, Ghilea et al. 2011, and Ruffolo et al. 2012). All these approaches are basically extensions of the
original NLGC theory.

A very different approach was proposed by Shalchi (2010), namely the Unified Non-Linear (UNLT) transport theory.
The main problem in analytical theories for perpendicular diffusion is the emergence of 4th order correlation functions.
In the NLGC theory and the aforementioned extensions, such 4th order correlations are approximated by a produced
of two 2nd order correlations. The 2nd order correlations are then approximated by different models such as a
diffusion model (Matthaeus et al. 2003, Shalchi 2006) or a random ballistic model (Ghilea et al. 2011, Ruffolo et al. 2012).
The UNLT theory is based on the direct evaluation of 4th order correlations by using the (pitch-angle dependent) Fokker-Planck
equation. The UNLT theory correctly describes subdiffusive transport in slab turbulence and contains the correct
FLRW limit without specifying the turbulence properties\footnote{The UNLT theory contains the correct FLRW limit
and the Matthaeus et al. (1995) theory. The NLGC theory does not contain this limit. However, it was shown before (see, e.g., Minnie et al. 2009)
that for two-dimensional turbulence and certain forms of the spectrum, the FLRW limit can be obtained.} (see, e.g., Shalchi 2014).
It also contains the NLGC theory and the field line diffusion theory of Matthaeus et al. (1995) as special limits
justifying the name Unified Non-Linear transport theory.

In the current paper we compute the perpendicular diffusion coefficient based on the NLGC theory and the UNLT theory.
In the following two paragraphs these two theories are discussed.
\subsection{The Non-Linear Guiding Center Theory}
In Matthaeus et al. (2003) the so-called Non-Linear Guiding Center (NLGC) theory was derived. The latter theory
is based on several assumptions leading to the following non-linear integral equation for the perpendicular diffusion
coefficient
\be
\kappa_{\perp} = \frac{a^2 v^2}{3 B_0^2} \int d^3 k \; \frac{P_{xx} (\vec{k})}{\kappa_{\parallel} k_{\parallel}^2 + \kappa_{\perp} k_{\perp}^2 + v/\lambda_{\parallel}}.
\label{NLGC}
\ee
Here we used the wavevector $\vec{k}$, the magnetic correlation tensor
\be
P_{mn} \left( \vec{k} \right) = \left< \delta B_m \left( \vec{k} \right) \delta B_n^{*} \left( \vec{k} \right) \right>,
\label{BBcorr}
\ee
the parallel diffusion coefficient of the particle $\kappa_{\parallel}$, the parallel mean free path $\lambda_{\parallel} = 3 \kappa_{\parallel} /v$,
the mean magnetic field $B_0$, and the particle speed $v$. We have also used the parameter $a^2$ which is related to the probability that the particle
is tied to a single magnetic field line. Eq. (\ref{NLGC}) was derived under the assumption that $\delta B_z \ll B_0$ and that the turbulence is static.
\subsection{The UNLT theory}
Because the NLGC theory is problematic in some cases, Shalchi (2010) derived the so-called Unified Non-Linear Transport (UNLT) theory.
The latter theory still provides a nonlinear integral equation for the perpendicular diffusion coefficient like the NLGC theory. However, the
theory contains different terms in the denominator\footnote{Eqs. (\ref{NLGC}) and (\ref{UNLT}) look very similar. However, the term (\ref{defAAA})
in the UNLT theory is completely different compared to the corresponding term in NLGC theory. Therefore, we expect that at least in certain limits,
the two theories provide very different solutions.}
\be
\kappa_{\perp} = \frac{a^2 v^2}{3 B_0^2} \int d^3 k \; \frac{P_{xx} (\vec{k})}{F(k_{\parallel},k_{\perp}) + (4/3) \kappa_{\perp} k_{\perp}^2 + v/\lambda_{\parallel}}
\label{UNLT}
\ee
where we have used
\be
F(k_{\parallel},k_{\perp}) = \frac{(2 v k_{\parallel} /3)^2}{(4/3) \kappa_{\perp} k_{\perp}^2} \equiv \frac{v^2 k_{\parallel}^2}{3 \kappa_{\perp} k_{\perp}^2}.
\label{defAAA}
\ee
The parameters used here are the same as in Eq. (\ref{NLGC}). Although the integral equation (\ref{UNLT}) has some similarities with Eq. (\ref{NLGC}),
the two theories provide different results in the general case (see Tautz \& Shalchi 2011).
\section{Noisy Reduced Magnetohydrodynamic Turbulence}
Ruffolo \& Matthaeus (2013) proposed the NRMHD turbulence model. All details can be found in the aforementioned
paper. In the following we discuss some aspects of this model and its relation to two-dimensional turbulence.
\subsection{The correlation tensor for NRMHD turbulence}
In the following we discuss the magnetic correlation tensor (\ref{BBcorr}) for the NRMHD model. According to Ruffolo \& Matthaeus (2013),
the two relevant components of the magnetic correlation tensor have the form
\bdm
P_{xx} \left( \vec{k} \right) = \frac{1}{4 \pi K} k_y^2 A \left( k_{\perp} \right)
\left\{
\begin{array}{ccc}
1 & \quad \textnormal{if} & \quad \left| k_{\parallel} \right| \leq K \\
0 & \quad \textnormal{if} & \quad \left| k_{\parallel} \right| > K.
\end{array}
\right.
\label{Pxx}
\edm
and
\bdm
P_{yy} \left( \vec{k} \right) = \frac{1}{4 \pi K} k_x^2 A \left( k_{\perp} \right)
\left\{
\begin{array}{ccc}
1 & \quad \textnormal{if} & \quad \left| k_{\parallel} \right| \leq K \\
0 & \quad \textnormal{if} & \quad \left| k_{\parallel} \right| > K.
\end{array}
\right.
\label{Pyy}
\edm
In the model described here, we used the (axi-symmetric) spectrum $A (k_{\perp})$ and the parameter $K$ which cuts off
the spectrum in the parallel direction. Compared to the tensor discussed in Ruffolo \& Matthaeus (2013), we have
different prefactors in our model because we use a different form of the Fourier-transform. As in Ruffolo \& Matthaeus (2013)
we only consider the special case of axi-symmetric turbulence where the spectrum depends only on $k_{\perp}$. We like
to emphasize that $\delta B_z =0$ in the model considered here and, therefore $P_{zz} =0$. Before we discuss the
spectrum $A \left( k_{\perp} \right)$ we briefly think about the normalization. Since we have to satisfy
\bdm
\delta B^2
& = & \delta B_x^2 + \delta B_y^2 \nonumber\\
& = & \int d^3 k \; \left[ P_{xx} \left( \vec{k} \right) + P_{yy} \left( \vec{k} \right) \right] \nonumber\\
& = & \int_{0}^{\infty} d k_{\perp} \; k_{\perp}^3 A \left( k_{\perp} \right)
\label{normal}
\edm
we can determine the constants in the spectrum $A \left( k_{\perp} \right)$.
\subsection{The spectrum $A \left( k_{\perp} \right)$}
A key element in theories for particle transport and field line random walk is the turbulence spectrum. Especially the
large scales (corresponding to small wavenumbers) of the spectrum control field line diffusion coefficients and perpendicular
diffusion coefficients of energetic particles (see, e.g., Shalchi \& Kourakis 2007, Shalchi \& Weinhorst 2009, Minnie et al. 2009,
Shalchi et al. 2010). In the following we use exactly the spectrum proposed by Ruffolo \& Matthaeus (2013) which has the form
\be
A \left( k_{\perp} \right) = \frac{A_0}{\left[ 1 + \left( k_{\perp} l_{\perp} \right)^2 \right]^{7/3}}.
\label{spec1}
\ee
Here we have used the characteristic length scale $l_{\perp}$ which denotes the turnover from the energy range of the spectrum
to the inertial range. Therefore, the latter scale is also known as the bendover scale. Usually this scale is directly proportional
to the integral scale of the turbulence (see, e.g., Shalchi 2014). By using the normalization condition (\ref{normal}), we can
specify the parameter $A_0$
\be
\delta B^2 = A_0 \int_{0}^{\infty} d k_{\perp} \; \frac{k_{\perp}^3}{\left[ 1 + \left( k_{\perp} l_{\perp} \right)^2 \right]^{7/3}}.
\ee
The latter integral can be solved (see, e. g., Gradshteyn \& Ryzhik 2000) and it yields $9/8$. Therefore, one can easily
determine the parameter $A_0$ and the spectrum (\ref{spec1}) becomes
\be
A \left( k_{\perp} \right) = \frac{8}{9} l_{\perp}^4 \delta B^2 \frac{1}{\left[ 1 + \left( k_{\perp} l_{\perp} \right)^2 \right]^{7/3}}.
\label{spec2}
\ee
With Eq. (\ref{Pxx}) we now know the $xx$-component of the correlation tensor which enters Eqs. (\ref{NLGC}) and (\ref{UNLT}).
Now our turbulence model is complete and in Sect. 4 we use it to compute the perpendicular diffusion coefficient.
\subsection{Relation to the two-dimensional model}
The NRMHD model can be seen as an extension/generalization of the pure two-dimensional model which was often used before in
the literature (see, e.g., Fyfe \& Montgomery 1976, Fyfe et al. 1977). The pure two-dimensional model is sometimes called
reduced MHD model (see, e.g., Strauss 1976, Montgomery 1982, and Higdon 1984). Some aspects of the corresponding spectrum
are discussed in Matthaeus et al. (2007) and Shalchi \& Weinhorst (2009).

In analytical treatments of turbulence, random walking magnetic field lines, and perpendicular transport of energetic
particles, physical quantities are usually given as wavenumber integral. Let's assume that we have an analytical
theory for the quantity $\xi_{xx}$ given as
\be
\xi_{xx} = \int d^3 k \; P_{xx} \left( \vec{k} \right) \chi \left( k_{\parallel}, k_{\perp} \right).
\label{genform}
\ee
Examples are the NLGC theory (\ref{NLGC}), the UNLT theory (\ref{UNLT}), and the normalization condition (\ref{normal}).
Now we evaluate the latter form by using the NRMHD model (\ref{Pxx}). In this case we find
\be
\xi_{xx} (K)
= \frac{1}{4 K} \int_{-K}^{+K} d k_{\parallel} \int_{0}^{\infty} d k_{\perp} \;
k_{\perp}^3 A (k_{\perp}) \chi \left( k_{\parallel}, k_{\perp} \right).
\ee
Now we consider the limit
\be
\xi_{xx}^{2D} := \lim_{K \rightarrow 0} \xi_{xx} (K)
\ee
and we obtain
\be
\xi_{xx}^{2D} = \frac{1}{2} \int_{0}^{\infty} d k_{\perp} \; k_{\perp}^3 A (k_{\perp}) \chi \left( k_{\parallel}=0, k_{\perp} \right).
\label{xilimit}
\ee
The pure two-dimensional model is defined as
\be
P_{xx} \left( \vec{k} \right) = g^{2D} (k_{\perp}) \frac{\delta (k_{\parallel})}{k_{\perp}} \frac{k_{y}^2}{k_{\perp}^2}.
\ee
Using this model with the form (\ref{genform}) we obtain
\be
\xi_{xx}^{2D} = \pi \int_{0}^{\infty} d k_{\perp} \; g^{2D} (k_{\perp}) \chi \left( k_{\parallel}=0, k_{\perp} \right).
\label{2dform}
\ee
The latter form can be compared with Eq. (\ref{xilimit}) to find the correspondence
\be
g^{2D} (k_{\perp}) = \frac{1}{2 \pi} k_{\perp}^3 A (k_{\perp}).
\label{specrelation}
\ee
Obviously the spectrum $A (k_{\perp})$ is directly related to the spectrum used in the pure two-dimensional turbulence
model $g^{2D} (k_{\perp})$. By combining the latter relation with the spectrum (\ref{spec2}), one can easily show that
this spectrum is a special case of the Shalchi \& Weinhorst (2009) model if we set $s=5/3$ and $q=3$ therein.
\section{Computing the Perpendicular Diffusion Coefficient}
In the following, we compute the perpendicular diffusion coefficient based on the two theories discussed in Sect. 2.
Eqs. (\ref{NLGC}) and (\ref{UNLT}) have the form
\be
\kappa_{\perp} = \frac{a^2 v^2}{3 B_0^2} \int d^3 k \; \frac{P_{xx} (\vec{k})}{U (k_{\perp}) + V (k_{\perp}) \; k_{\parallel}^2}
\label{kappa1}
\ee
where the functions $U (k_{\perp})$ and $V (k_{\perp})$ are different for the two considered theories. They are summarized
in Table \ref{tabUV}. With Eq. (\ref{Pxx}) this becomes
\bdm
\kappa_{\perp} & = & \frac{a^2 v^2}{3 B_0^2} \frac{1}{2 K} \nonumber\\
& \times & \int_{0}^{\infty} d k_{\perp} \; k_{\perp}^3 A \left( k_{\perp} \right) \frac{\arctan \left( K \sqrt{V/U} \right)}{\sqrt{U V}}.
\label{kappa2}
\edm
Here we kept the spectrum $A \left( k_{\perp} \right)$ in the equation for the perpendicular diffusion coefficient. Later we will replace
it by the form (\ref{spec2}). Very easily one can consider the limit $K \rightarrow 0$ and by using $\arctan (x) \approx x$, one can
derive the corresponding integral equation for two-dimensional turbulence from (\ref{kappa2}).

\begin{table}[t]
\caption{The functions $U (k_{\perp})$ and $V (k_{\perp})$ for NLGC and UNLT theories.}
\begin{center}
\begin{tabular}{lll}\hline
Parameter 		& NLGC theory												& UNLT theory													\\
\hline
$U (k_{\perp})$	& $ \kappa_{\perp} k_{\perp}^2 + v/\lambda_{\parallel} $ 	& $ (4/3) \kappa_{\perp} k_{\perp}^2 + v/\lambda_{\parallel} $ 	\\
$V (k_{\perp})$	& $ \kappa_{\parallel} $ 									& $ v^2 / (3 \kappa_{\perp} k_{\perp}^2) $ 						\\
\hline
\end{tabular}
\end{center}
\medskip
\label{tabUV}
\end{table}
\subsection{The FLRW Limit from UNLT Theory}
\begin{figure}
\centering 
\includegraphics[width=0.48\textwidth]{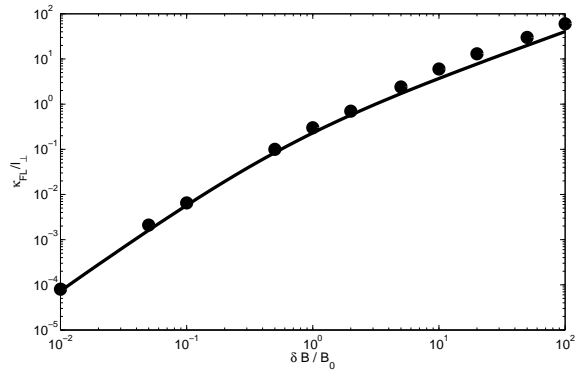}
\caption{The field line diffusion coefficient for NRMHD turbulence. Shown is $\kappa_{FL} / l_{\perp}$ versus the magnetic field
ratio $\delta B / B_0$ for $\tilde{K} \equiv K l_{\perp}= 1$. The solid line represents the analytical result obtained by solving
Eq. (\ref{fieldline2}) numerically and the dots represent the simulations (see Sect. 5 for details). We like to emphasize that
the solid line is in agreement with the result obtained by Ruffolo \& Matthaeus (2013).}
\label{fieldlinediffusion}
\end{figure}

One strength of the UNLT theory is that the correct Field Line Random Walk (FLRW) limit can be derived from the theory. In the present paragraph
we demonstrate this for the turbulence model considered here. We can obtain the FLRW limit by suppressing parallel diffusion and by forcing the
particle to follow magnetic field lines. This means that we have to set $a^2 = 1$ and $v/\lambda_{\parallel} = 0$ in Eq. (\ref{kappa2}) and in
the functions $U$ and $V$ listed in Table \ref{tabUV}. Therefore, we have $U = (4/3) \kappa_{\perp} k_{\perp}^2$ and, thus, $\sqrt{U \; V} = 2v/3$
and $\sqrt{V/U} = v/(2 \kappa_{\perp} k_{\perp}^2)$. With these relations, Eq. (\ref{kappa2}) becomes
\be
\kappa_{\perp} = \frac{v}{4 K B_0^2}
\int_{0}^{\infty} d k_{\perp} \; k_{\perp}^3 A \left( k_{\perp} \right) \arctan \left( \frac{v K}{2 \kappa_{\perp} k_{\perp}^2} \right).
\label{fieldline1}
\ee
The latter equation has the solution
\be
\kappa_{\perp} = \frac{v}{2} \kappa_{FL},
\label{flrwlimit1}
\ee
or, in terms of mean free paths
\be
\lambda_{\perp} = \frac{3}{2} \kappa_{FL}
\label{flrwlimit2}
\ee
with the field line diffusion coefficient
\be
\kappa_{FL} = \frac{1}{2 K B_0^2}
\int_{0}^{\infty} d k_{\perp} \; k_{\perp}^3 A \left( k_{\perp} \right) \arctan \left( \frac{K}{\kappa_{FL} k_{\perp}^2} \right).
\label{fieldline2}
\ee
Eqs. (\ref{flrwlimit1}) and (\ref{flrwlimit2}) correspond to the FLRW limit and Eq. (\ref{fieldline2}) agrees perfectly
with Eq. (25) of Ruffolo \& Matthaeus (2013). We like to emphasize that Eq. (\ref{fieldline2}) was derived from Eq. (\ref{UNLT})
representing the UNLT theory by setting $a^2 =1$ and $v/\lambda_{\parallel}=0$ therein. In Ruffolo \& Matthaeus (2013)
the same result was obtained by employing the field line diffusion theory of Matthaeus et al. (1995). As shown here the UNLT theory of
Shalchi (2010) allows to describe perpendicular diffusion of energetic particles as well as the diffusion of magnetic field lines.
In Fig. \ref{fieldlinediffusion} we show the numerical solution of Eq. (\ref{fieldline2}). In the latter figure we also show simulations
of FLRW confirming the validity of Eq. (\ref{fieldline2}). More details about this numerical work can be found in Sect. 5.

\subsection{The limit $K \rightarrow 0$}
The UNLT theory represented by Eq. (\ref{UNLT}) is a special case of the form (\ref{genform})
with $\xi_{xx} = \kappa_{xx} = \kappa_{\perp}$ and
\be
\chi \left( k_{\parallel}, k_{\perp} \right) = \frac{a^2 v^2}{3 B_0^2} \frac{1}{F(k_{\parallel}, k_{\perp}) + (4/3) \kappa_{\perp} k_{\perp}^2 + v/\lambda_{\parallel}}
\ee
with the function $F(k_{\parallel}, k_{\perp})$ defined in (\ref{defAAA}). According to Eq. (\ref{2dform}) this becomes in the limit $K \rightarrow 0$
\be
\kappa_{\perp} = \frac{\pi}{3} \frac{a^2 v^2}{B_0^2} \int_{0}^{\infty} d k_{\perp} \;
\frac{g^{2D} (k_{\perp})}{(4/3) \kappa_{\perp} k_{\perp}^2 + v/\lambda_{\parallel}}
\ee
corresponding to Eq. (4) of Shalchi (2013b). The spectrum $g^{2D} (k_{\perp})$ is related to the $A (k_{\perp})$ via Eq. (\ref{specrelation})
of the present paper. Therefore, in the limit $K \rightarrow 0$, we expect to find the results derived earlier for two-dimensional turbulence.
We like to emphasize that strictly pure two-dimensional turbulence should be considered to be a singular case and that diffusion
theories such as NLGC and UNLT theories are not longer valid for that specific model of turbulence.
\subsection{Perpendicular diffusion for the general case}
Here we go back to the general form (\ref{kappa2}) with the functions $U (k_{\perp})$ and $V (k_{\perp})$ from
Table \ref{tabUV}. Eq. (\ref{kappa2}) has to be evaluated numerically. Therefore, we introduce new quantities which
are more appropriate for numerical treatments of the transport. In the following we use
\bdm
\tilde{K} & = & l_{\perp} K, \nonumber\\
S & = & K \sqrt{V / U}, \nonumber\\
Q & = & \frac{1}{v} \sqrt{U \; V},
\edm
and instead of using the spatial diffusion coefficient we use mean free paths defined as $\lambda_{\parallel} = 3 \kappa_{\parallel} /v$
and $\lambda_{\perp} = 3 \kappa_{\perp} /v$. By using the latter parameters, Eq. (\ref{kappa2}) becomes 
\be
\frac{\lambda_{\perp}}{l_{\perp}} = \frac{a^2}{2 \tilde{K} B_0^2}
\int_{0}^{\infty} d k_{\perp} \; k_{\perp}^3 A \left( k_{\perp} \right) \frac{\arctan \left[ S (k_{\perp}) \right]}{Q (k_{\perp})}
\ee
where the parameters/functions $S$ and $Q$ can be found below. To proceed we employ the spectrum (\ref{spec2})
and we use the integral transformation $x = l_{\perp} k_{\perp}$ to obtain
\be
\frac{\lambda_{\perp}}{l_{\perp}} = \frac{4 a^2}{9 \tilde{K}} \frac{\delta B^2}{B_0^2}
\int_{0}^{\infty} d x \; \frac{x^3}{\left( 1 + x^2 \right)^{7/3}} \frac{\arctan \left[ S(x) \right]}{Q(x)}.
\label{finaleq}
\ee
The two functions $S(x)$ and $Q(x)$ are different for NLGC and UNLT theories. For the NLGC theory we have to use
\be
S_N (x) = \tilde{K} \sqrt{ \left( \frac{\lambda_{\parallel}}{3 l_{\perp}} \right) / \left( \frac{\lambda_{\perp}}{3 l_{\perp}} x^2 + \frac{l_{\perp}}{\lambda_{\parallel}} \right) }
\label{Rnlgc}
\ee
and
\be
Q_N (x) = \sqrt{ \left( \frac{\lambda_{\perp}}{3 l_{\perp}} x^2 + \frac{l_{\perp}}{\lambda_{\parallel}} \right) \frac{\lambda_{\parallel}}{3 l_{\perp}}}.
\label{Qnlgc}
\ee
For the UNLT theory, however, we have
\be
S_U (x) = \tilde{K} \sqrt{ \left( \frac{l_{\perp}}{\lambda_{\perp} x^2} \right) / \left( \frac{4 \lambda_{\perp}}{9 l_{\perp}} x^2 + \frac{l_{\perp}}{\lambda_{\parallel}} \right) }
\label{Runlt}
\ee
and
\be
Q_U (x) = \sqrt{ \frac{4}{9} + \frac{l_{\perp}^2}{\lambda_{\parallel} \lambda_{\perp} x^2}}.
\label{Qunlt}
\ee
In the following we compute the perpendicular mean free path versus the parallel mean free path for different values of the
parameters $a^2$, $\delta B^2 / B_0^2$, and $\tilde{K}$. The used values are listed in the caption of the corresponding figure.
By specifying these parameters we can solve Eq. (\ref{finaleq}) for the NLGC theory and the UNLT theory numerically.

\begin{figure}
\centering 
\includegraphics[width=0.48\textwidth]{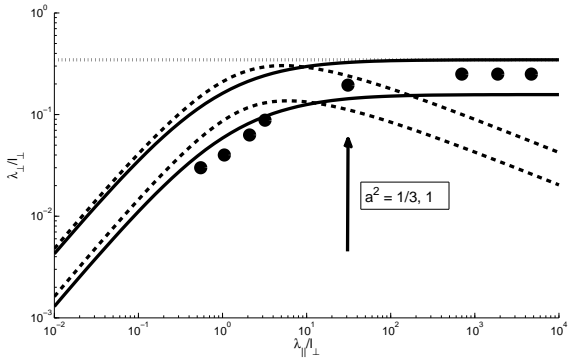}
\caption{The perpendicular mean free path versus the parallel mean free path for the NRMHD model. We compare the NLGC theory (dashed line)
with the UNLT theory (solid line), and the field line random walk limit (dotted line). Here we set $\tilde{K} = 1$ and $\delta B^2 / B_0^2 = 1$.
The dots represent the test-particle simulations discussed in Section 5. The two analytical theories were evaluated for two different values
of the parameter $a$.}
\label{lperpvslpara}
\end{figure}

In Fig. \ref{lperpvslpara} we compute the perpendicular mean free path versus the parallel mean free path for two different
values of the parameter $a^2$ and $\tilde{K}=1$ and $\delta B^2 / B_0^2 =1$. Shown are also test-particle simulations which are
discussed in Section 5. We can easily see that for NRMHD turbulence, there are two regimes. In the regime $\lambda_{\parallel} \ll l_{\perp}$
the perpendicular mean free path increases linearly with the parallel mean free path. In this regime NLGC and UNLT theories
provide very similar results. Below we will discuss that this similarity cannot be found in the general case. As soon as the parallel
mean free path becomes longer than the bendover scale $l_{\perp}$, the two theories provide very different results. Whereas
the perpendicular mean free path obtained from NLGC theory decreases with increasing $\lambda_{\parallel}$, the UNLT provides
a perpendicular mean free path which becomes constant. In the case $\lambda_{\parallel} \gg l_{\perp}$, the results obtained
from UNLT theory are very close to the FLRW limit $\lambda_{\perp} = 3 \kappa_{FL} /2$. In Appendix A we consider the limit
$\lambda_{\parallel} \rightarrow \infty$ in NLGC theory. There it is shown that in this limit we find $\lambda_{\perp} \sim \lambda_{\parallel}^{-1/3}$
in disagreement with the UNLT theory and simulations\footnote{We like to point out that we indeed find the exponent
$-1/3$ for the dependence on the parallel mean free path. In Shalchi et al. (2004), for instance, it was derived
$\lambda_{\perp} \sim \lambda_{\parallel}^{+1/3}$ which is different compared to the result derived in the present
paper. The exponent was $+1/3$ derived for pure two-dimensional turbulence and a very specific spectrum. Therefore,
this results has nothing to do with the exponent we derived in the current paper.}.

\begin{figure}
\centering 
\includegraphics[width=0.48\textwidth]{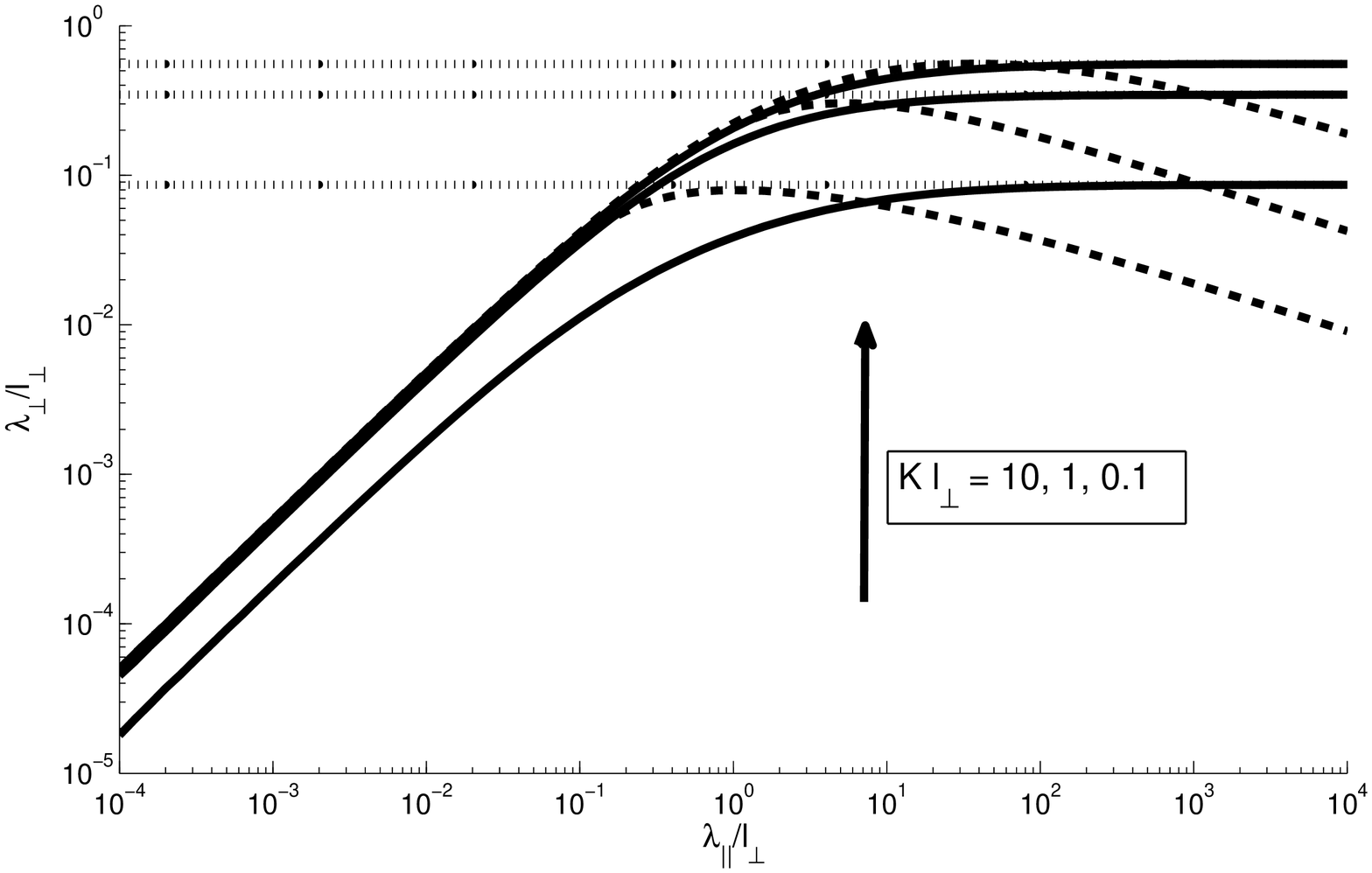}
\caption{The perpendicular mean free path versus the parallel mean free path for the NRMHD model. We compare the NLGC theory (dashed line)
with the UNLT theory (solid line), and the field line random walk limit (dotted line). We compute $\lambda_{\perp}$ for $\tilde{K} = K l_{\perp} = 10, 1, 0.1$.
Here we set $a^2 = 1$ and $\delta B^2 / B_0^2 = 1$.}
\label{lperpvslparavarK}
\end{figure}

\begin{figure}
\centering 
\includegraphics[width=0.48\textwidth]{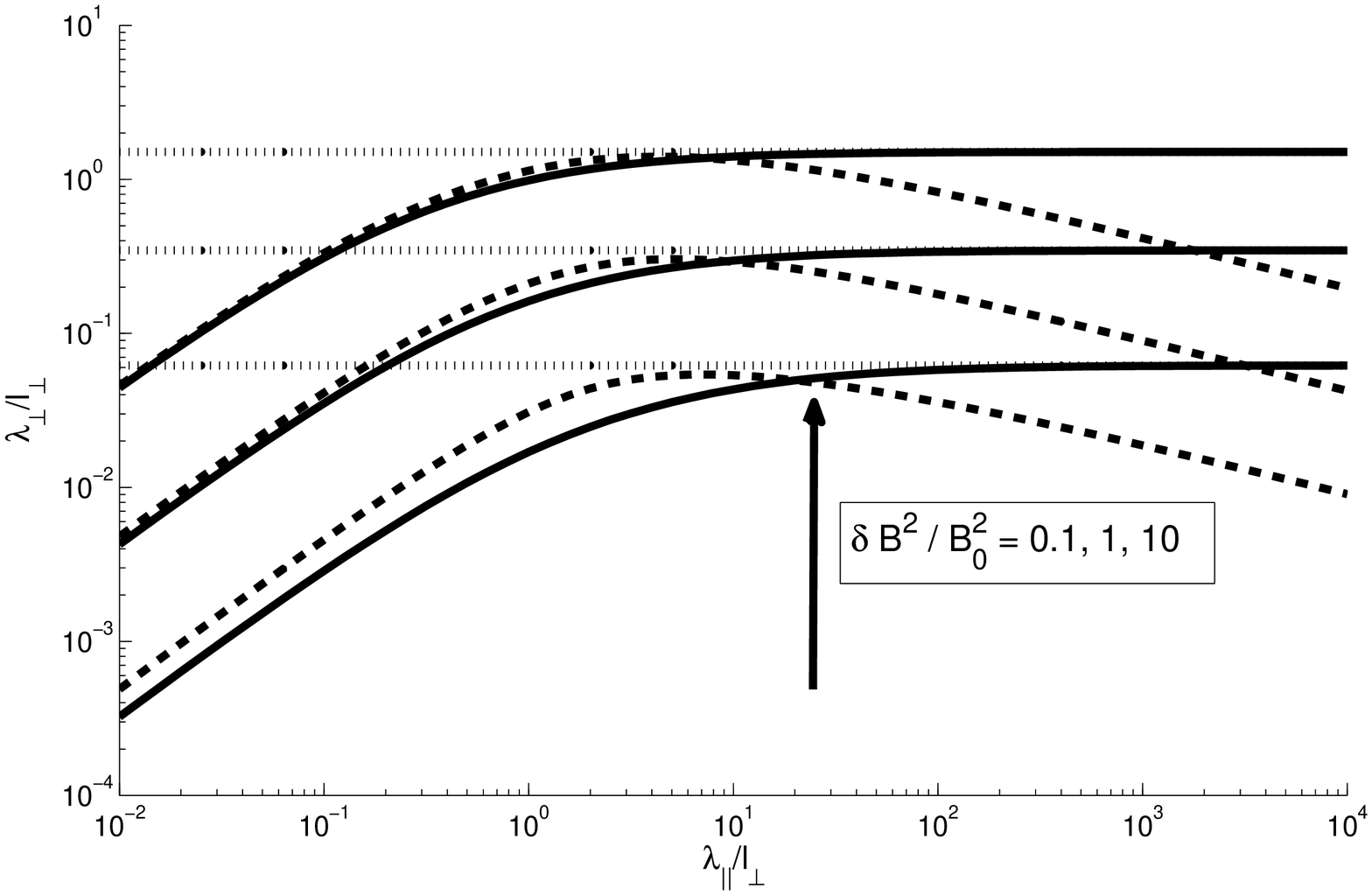}
\caption{The perpendicular mean free path versus the parallel mean free path for the NRMHD model. We compare the NLGC theory (dashed line)
with the UNLT theory (solid line), and the field line random walk limit (dotted line). We compute $\lambda_{\perp}$ for $\delta B^2 / B_0^2 = 0.1, 1, 10$.
Here we set $a^2 = 1$ and $\tilde{K} = 1$.}
\label{lperpvslparavardb}
\end{figure}

In Figs. \ref{lperpvslparavarK} and \ref{lperpvslparavardb} we study the influence of the two parameters $\delta B / B_0$
and $\tilde{K}$ onto the perpendicular diffusion coefficient. For small $\tilde{K} \rightarrow 0$ the perpendicular mean
free path approaches the results one would obtain for two-dimensional turbulence whereas for larger values of $\tilde{K}$
the perpendicular mean free path is getting shorter. For such large values of $\tilde{K}$, Fig. \ref{lperpvslparavarK}
also show a further discrepancy between NLGC and UNLT theories. For $\tilde{K} = 10$ the two theories disagree with
each other even if the parallel mean free path is very short. NLGC theories predicts that the ratio $\lambda_{\perp}/\lambda_{\parallel}$
does not depend on the parameter $\tilde{K}$ (see appendix A of the current paper) whereas UNLT theory clearly states a
dependence on this parameter. This discrepancy has to be subject of future work and therewith analytical solutions of
the UNLT integral equation for NRMHD turbulence. From Fig. \ref{lperpvslparavardb}, one can see that
the perpendicular mean free path depends sensitively on the magnetic field ratio $\delta B / B_0$. For weak turbulence
amplitudes such as $\delta B^2 / B_0^2 = 0.1$, we find again a discrepancy between NLGC and UNLT theories. Obviously,
these two theories provide different results for most turbulence and particle parameters.

We like to emphasize that all our results were obtained for a specific spectrum, namely the model spectrum given
by Eq. (\ref{spec2}) which is the spectrum proposed by Ruffolo \& Matthaeus (2013). For a different spectrum (e.g., a
different spectral index in the energy range, a spectrum with cut-off at small wavenumbers) one could obtain different
results and the differences between NLGC and UNLT theories could be smaller or larger in such cases.

\section{Simulations}
A powerful tool in order to test analytical theories such as NLGC or UNLT theories are test-particle simulations. In the current section
we use an extension of the code used in Hussein \& Shalchi (2014). In the following we discuss some technical details of that code, the
results obtained for the field line diffusion coefficient, and the simulated parallel and perpendicular diffusion coefficients.
\subsection{The test-particle code}
Test-particle simulations have been performed before. In Hussein \& Shalchi (2014), for instance, we have used a code to simulate the
interaction between energetic particles and different turbulence models. These models were the slab model, the isotropic model, and
a composition of slab and two-dimensional modes. In all these models only one independent wavevector component controls the turbulent
magnetic field. The NRMHD model considered here is more complicated because two components are relevant, namely $k_{\parallel}$ and
$k_{\perp}$. Therefore, one has to evaluate an extra sum numerically making the simulations much more time-consuming.
We describe the technical details of our numerical tool in Appendix B and focus on the results in the main part of the paper.
\subsection{The field line diffusion coefficient}
By solving the field line equation $dx = dz \delta B_x / B_0$ numerically, one can obtain the field line diffusion coefficient for
different parameter values. Here we set $\tilde{K} = 1$ and compute the field line diffusion coefficient for different values of the
magnetic field ratio $\delta B / B_0$. The results are listed in Table \ref{flsimtable} and they are compared with the analytical
results in Fig. \ref{fieldlinediffusion}. As shown in the latter figure, the agreement between analytical theory and simulations is
very good confirming the nonlinear theory for field line diffusion developed by Matthaeus et al. (1995) and the UNLT theory of
Shalchi (2010). Our simulations for random walking magnetic field lines agree well with the simulations presented in
Snodin et al. (2013).

\begin{table*}[t]
\caption{The simulated field line diffusion coefficient for the NRMHD model versus the ratio $\delta B / B_{0}$. For all simulation runs we set $\tilde{K}=1.0$.}
\begin{center}
\begin{tabular}{llllllllllll}
\hline
$\delta B / B_{0}$				& $0.01$				& $0.05$ 	& $0.1$		& $0.5$ 	& $1.0$		& $2.0$	& $5.0$	& $10.0$	& $20.0$	& $50.0$	& $100.0$	\\
\hline
$\kappa_{FL} / l_{\perp}$		& $8.0 \cdot 10^{-5}$	& $0.0021$ 	& $0.0065$	& $0.1$ 	& $0.3$		& $0.7$	& $2.4$	& $6.0$		& $13.0$	& $30.0$	& $60.0$	\\
\hline
\end{tabular}
\end{center}
\label{flsimtable}
\end{table*}

\subsection{The particle diffusion coefficients}
In the current paragraph we use the code described above to compute parallel and perpendicular mean free paths. For these simulations
we set $\tilde{K} = 1$ and $\delta B / B_0 = 1$. Our results are listed in Table \ref{simtable} and they are visualized in
Fig. \ref{lperpvslpara}. It is shown that the numerical perpendicular mean free path agrees well with NLGC and UNLT theories for
the case of small parallel mean free paths. For long parallel mean free paths, however, only the UNLT theory agrees with the simulations.
The decreasing perpendicular mean free path for larger values of $\lambda_{\parallel}$ provided by NLGC theory cannot be seen in the
simulations. The prediction of UNLT theory that the perpendicular mean free path approaches asymptotically the FLRW limit, in contrast,
can also be seen in the numerical work. Therefore, UNLT theory is confirmed once again.

\begin{figure}
\centering 
\includegraphics[width=0.48\textwidth]{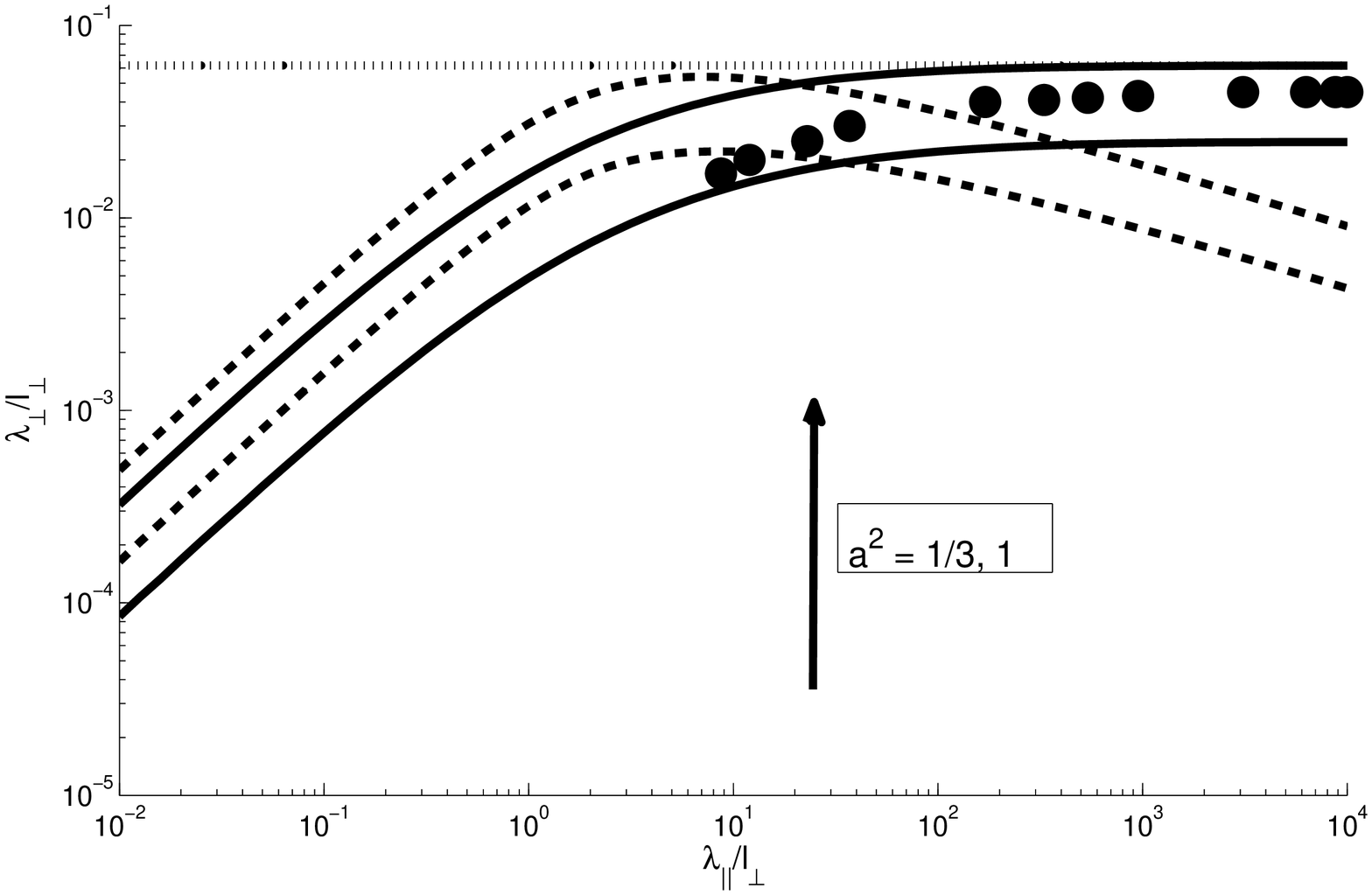}
\caption{The perpendicular mean free path versus the parallel mean free path for the NRMHD model. We compare the NLGC theory (dashed line)
with the UNLT theory (solid line), and the field line random walk limit (dotted line). Here we set $\tilde{K} = 1$ and $\delta B^2 / B_0^2 = 0.1$.
The dots represent the simulations.}
\label{lperpvslparaweak}
\end{figure}

\begin{figure}
\centering 
\includegraphics[width=0.48\textwidth]{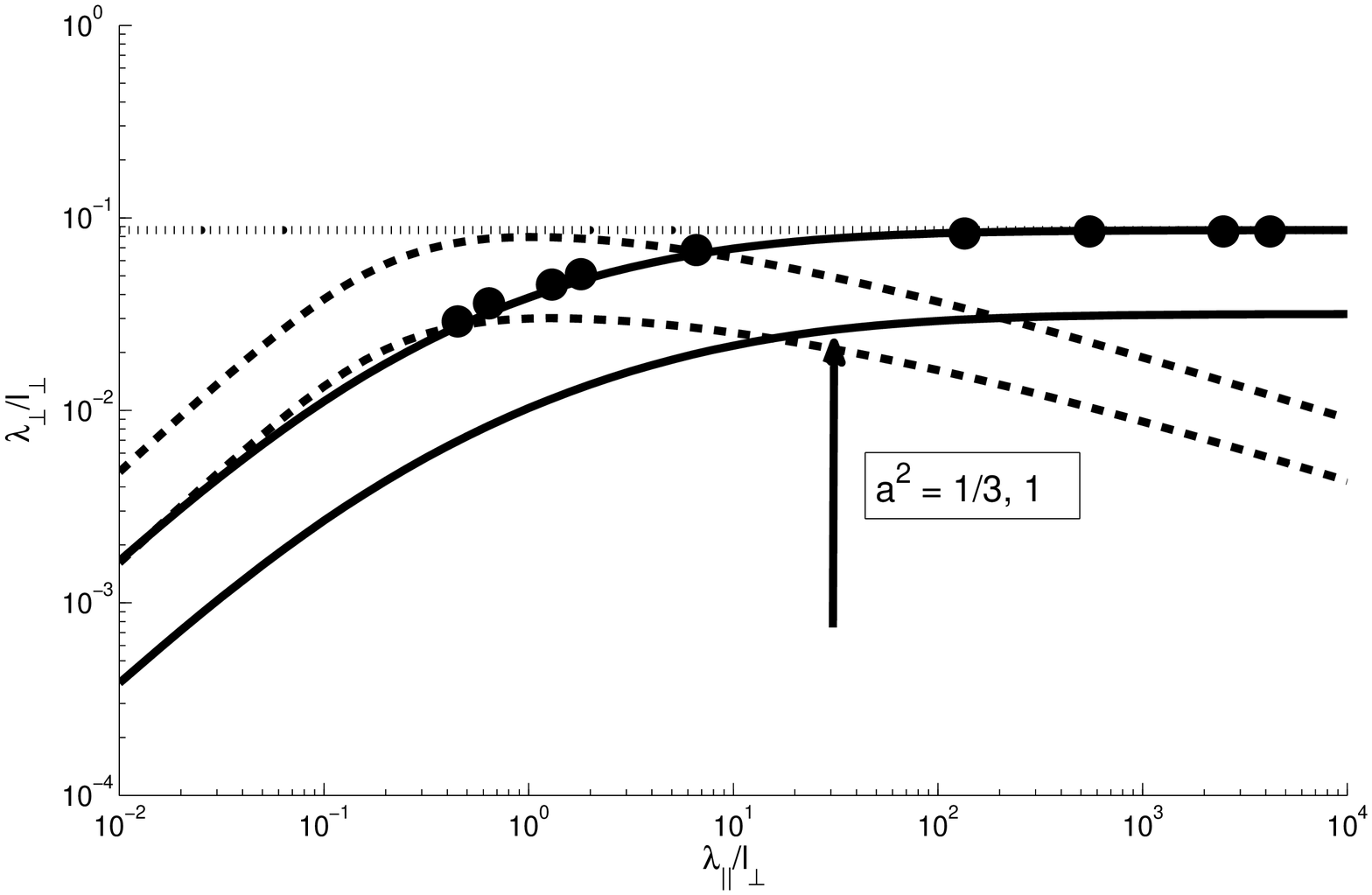}
\caption{The perpendicular mean free path versus the parallel mean free path for the NRMHD model. We compare the NLGC theory (dashed line)
with the UNLT theory (solid line), and the field line random walk limit (dotted line). Here we set $\tilde{K} = 10$ and $\delta B^2 / B_0^2 = 1.0$.
The dots represent the simulations.}
\label{lperpvslparaK10}
\end{figure}

In Figs. \ref{lperpvslparaweak} and \ref{lperpvslparaK10} we repeat the simulations for $\tilde{K} = 1$, $\delta B^2 / B_0^2 = 0.1$
and $\tilde{K} = 10$, $\delta B^2 / B_0^2 = 1.0$, respectively. The results are listed in Tables \ref{simtable2} and \ref{simtable3}.
Qualitatively the results are very similar compared to the previous run. We can see that now even for small values of $\lambda_{\parallel}$,
NLGC and UNLT theories disagree with each other. The simulations clearly support the UNLT theory. It seems, however, that the parameter $a^2$
depends on the values of $\tilde{K}$ and $\delta B^2 / B_0^2$. More investigations concerning the value of $a^2$ have to be done in the future.

\begin{table*}[t]
\caption{The simulated mean free paths along and across the mean magnetic field versus the dimensionless magnetic rigidity $R_L / l_{\perp}$.
Here we have used $\tilde{K} = 1.0$ and $\delta B^2 / B_0^2 = 1.0$.}
\begin{center}
\begin{tabular}{llllllllll}
\hline
$R_L / l_{\perp}$						& $0.001$	& $0.01$ 	& $0.05$ 	& $0.1$ 	& $1.0$		& $5.0$					& $10.0$				& $16.0$				& $20.0$				\\
\hline
$\lambda_{\parallel} / l_{\perp}$		& $0.55$	& $1.05$ 	& $2.1$ 	& $3.2$ 	& $31$		& $700$					& $1875$				& $4700$				& $1.1 \cdot 10^4$		\\
$\lambda_{\perp} / l_{\perp}$			& $0.03$	& $0.04$ 	& $0.063$ 	& $0.088$ 	& $0.195$	& $0.25$				& $0.25$				& $0.25$				& $0.25$				\\
\hline
\end{tabular}
\end{center}
\label{simtable}
\end{table*}

\begin{table*}[t]
\caption{The simulated mean free paths along and across the mean magnetic field versus the dimensionless magnetic rigidity $R_L / l_{\perp}$.
Here we have used $\tilde{K} = 10$ and $\delta B^2 / B_0^2 = 1.0$.}
\begin{center}
\begin{tabular}{llllllllll}
\hline
$R_L / l_{\perp}$						& $0.005$	& $0.01$ 	& $0.05$ 	& $0.1$ 	& $1.0$		& $5.0$					& $10.0$				& $20.0$				& $25.0$				\\
\hline
$\lambda_{\parallel} / l_{\perp}$		& $0.45$	& $0.64$	& $1.3$ 	& $1.8$ 	& $6.6$		& $135$					& $550$					& $2480$				& $4200$				\\
$\lambda_{\perp} / l_{\perp}$			& $0.029$	& $0.036$	& $0.045$ 	& $0.051$ 	& $0.068$	& $0.083$				& $0.085$				& $0.085$				& $0.085$				\\
\hline
\end{tabular}
\end{center}
\label{simtable2}
\end{table*}

\begin{table*}[t]
\caption{The simulated mean free paths along and across the mean magnetic field versus the dimensionless magnetic rigidity $R_L / l_{\perp}$.
Here we have used $\tilde{K} = 1.0$ and $\delta B^2 / B_0^2 = 0.1$.}
\begin{center}
\begin{tabular}{lllllllllllll}
\hline
$R_L / l_{\perp}$						& $0.005$	& $0.01$ 	& $0.05$ 	& $0.1$ 				& $1.0$					& $2.0$					& $3.0$					& $5.0$					& $10.0$				& $20.0$	& $25.0$						& $30.0$			\\
\hline
$\lambda_{\parallel} / l_{\perp}$		& $8.7$		& $12$	 	& $23$ 		& $37$	 				& $170$					& $330$					& $540$					& $950$					& $3100$				& $6300$	& $8750$						& $10^4$			\\
$\lambda_{\perp} / l_{\perp}$			& $0.017$	& $0.02$	& $0.025$ 	& $0.03$ 				& $0.04$				& $0.041$				& $0.042$				& $0.043$				& $0.045$				& $0.045$	& $0.045$						& $0.045$				\\
\hline
\end{tabular}
\end{center}
\label{simtable3}
\end{table*}

\section{Summary and Conclusion}
In Ruffolo \& Matthaeus (2013) the model of {\it Noisy Reduced MagnetoHydroDynamic (NRMHD) turbulence} was proposed and used to
compute the diffusion coefficient of random walking magnetic field lines based on the nonlinear diffusion theory of Matthaeus et al. (1995).
In the current paper we have investigated perpendicular diffusion of energetic particles by using two analytical theories, namely,
the {\it Non-Linear Guiding Center (NLGC) theory} of Matthaeus et al. (2003) and the {\it Unified Non-Linear Transport (UNLT)} theory
of Shalchi (2010). Furthermore we have performed test-particle simulations to obtain field line diffusion and particle transport
coefficients. We obtained the following results:
\begin{enumerate}
\item We have shown that the field line random walk limit with the correct field line diffusion coefficient can be obtained
from UNLT theory in the appropriate limit. For the case of NRMHD turbulence the field line diffusion coefficient
already obtained by Ruffolo \& Matthaeus (2013) is derived from UNLT theory. Our test-particle simulations confirm these
previous results and therewith our current understanding of field line diffusion (see Fig. \ref{fieldlinediffusion}).
\item The first time we obtain the perpendicular diffusion coefficient of energetic particles for NRMHD turbulence. We have
shown how the two parameters $\tilde{K} = K l_{\perp}$ and $\delta B / B_0$ influence the perpendicular mean free path.
The UNLT and NLGC theories provide very different results for the turbulence model considered here (see Figs. \ref{lperpvslpara},
\ref{lperpvslparaweak}, and \ref{lperpvslparaK10} of the current paper). According to the UNLT theory the perpendicular mean
free path increases linearly with the parallel mean free path $\lambda_{\parallel}$ and in the limit of large $\lambda_{\parallel}$
it becomes independent of the latter parameter. This behavior was already found for other turbulence models and agrees with
the universality of the transport discussed in detail in Shalchi (2014).
\item We have tested the validity of NLGC and UNLT theories by comparing them with test-particle simulations. As shown
in Figs. \ref{lperpvslpara}, \ref{lperpvslparaweak}, and \ref{lperpvslparaK10}, only the UNLT theory agrees with the simulations
for NRMHD turbulence. The scaling $\lambda_{\perp} \sim \lambda_{\parallel}^{-1/3}$ predicted by NLGC theory for long parallel
mean free paths cannot be seen in the simulations. Furthermore, we also find that for short parallel mean free path, 
NLGC theory predicts that the ratio $\lambda_{\perp}/\lambda_{\parallel}$ does not depend on the parameter $\tilde{K}$
whereas UNLT show a clear dependence.
\end{enumerate}

The UNLT theory originally developed by Shalchi (2010) can correctly describe field line diffusion and perpendicular transport
of energetic particles in NRMHD turbulence. This work, therefore, also complements previous work in which it has been shown
that UNLT theory can describe transport in two-component turbulence and Goldreich-Sridhar turbulence accurately (see, e.g.,
Tautz \& Shalchi 2011 and Shalchi 2013a). It will be subject of future work to derive analytical forms for the
perpendicular diffusion coefficient in NRMHD turbulence based on the UNLT theory.

\begin{acknowledgements}
{\it
M. Hussein and A. Shalchi acknowledge support by the Natural Sciences and Engineering Research Council (NSERC)
of Canada and national computational facility provided by WestGrid. We are also grateful to S. Safi-Harb for
providing her CFI-funded computational facilities for code tests and for some of the simulation runs presented here.
}
\end{acknowledgements}
\appendix
\section{A: Asymptotic limits derived from the NLGC integral equation}
Here we explore asymptotic limits one can obtain from the NLGC theory. A more detailed discussion of such
limits and the corresponding limits obtained from UNLT theory can be found in the main part of the text.
\subsection{The limit $\lambda_{\parallel} \rightarrow 0$}
Here we consider the (formal) limit $\lambda_{\parallel} \rightarrow 0$ in the NLGC integral equation. In this limit
Eqs. (\ref{Rnlgc}) and (\ref{Qnlgc}) provide
\bdm
S (x) & \rightarrow & \frac{\tilde{K} \lambda_{\parallel}}{\sqrt{3} l_{\perp}} \rightarrow 0 \nonumber\\
Q (x) & \rightarrow & \frac{1}{\sqrt{3}}
\edm
and, therefore,
\be
\arctan \left( S \right) \rightarrow S.
\ee
With the latter three limits, Eq. (\ref{finaleq}) becomes
\be
\frac{\lambda_{\perp}}{l_{\perp}} = \frac{4 a^2}{9 \tilde{K}} \frac{\delta B^2}{B_0^2} \frac{\tilde{K} \lambda_{\parallel}}{l_{\perp}}
\int_{0}^{\infty} d x \; \frac{x^3}{\left( 1 + x^2 \right)^{7/3}}.
\label{limit1b}
\ee
The $x$-integral can be solved by (see, e.g., Gradshteyn \& Ryzhik 2000)
\be
\int_{0}^{\infty} d x \; \frac{x^3}{\left( 1 + x^2 \right)^{7/3}} = \frac{9}{8}.
\ee
Therewith Eq. (\ref{limit1b}) becomes
\be
\frac{\lambda_{\perp}}{\lambda_{\parallel}} = \frac{a^2}{2} \frac{\delta B^2}{B_0^2}
\ee
which was derived before for two-dimensional turbulence (see, e.g., Shalchi et al. 2004).
\subsection{The limit $\lambda_{\parallel} \rightarrow \infty$}
Here we investigate the limit $\lambda_{\parallel} \rightarrow \infty$ in the NLGC integral equation. In this limit
Eqs. (\ref{Rnlgc}) and (\ref{Qnlgc}) provide
\bdm
S (x) & \rightarrow & \frac{\tilde{K}}{x} \sqrt{\frac{\lambda_{\parallel}}{\lambda_{\perp}}} \rightarrow \infty \nonumber\\
Q (x) & \rightarrow & \frac{\sqrt{\lambda_{\parallel} \lambda_{\perp}}}{3 l_{\perp}} x
\edm
and, therefore,
\be
\arctan \left( S \right) \rightarrow \pi / 2.
\ee
With the latter three limits, Eq. (\ref{finaleq}) becomes
\be
\frac{\lambda_{\perp}}{l_{\perp}} = \frac{2 \pi a^2}{3 \tilde{K}} \frac{\delta B^2}{B_0^2} \frac{l_{\perp}}{\sqrt{\lambda_{\parallel} \lambda_{\perp}}}
\int_{0}^{\infty} d x \; \frac{x^2}{\left( 1 + x^2 \right)^{7/3}}.
\label{limit1}
\ee
The $x$-integral can be solved by (see, e.g., Gradshteyn \& Ryzhik 2000)
\be
\int_{0}^{\infty} d x \; \frac{x^2}{\left( 1 + x^2 \right)^{7/3}} = \frac{\sqrt{\pi}}{4} \frac{\Gamma \left( 5/6 \right)}{\Gamma \left( 7/3 \right)}
\ee
where we have used the {\it Gamma function} $\Gamma (z)$. Therewith Eq. (\ref{limit1}) becomes
\be
\frac{\lambda_{\perp}}{l_{\perp}} = \frac{\pi^{3/2} a^2}{6 \tilde{K}} \frac{\Gamma \left( 5/6 \right)}{\Gamma \left( 7/3 \right)}
\frac{\delta B^2}{B_0^2} \frac{l_{\perp}}{\sqrt{\lambda_{\parallel} \lambda_{\perp}}}.
\label{limit2}
\ee
The latter equations can easily be solved by
\be
\frac{\lambda_{\perp}}{l_{\perp}} = \left[ \frac{\pi^{3/2} a^2}{6 \tilde{K}} \frac{\Gamma \left( 5/6 \right)}{\Gamma \left( 7/3 \right)}
\frac{\delta B^2}{B_0^2} \right]^{2/3} \left( \frac{l_{\perp}}{\lambda_{\parallel}} \right)^{1/3}.
\label{limit3}
\ee
A more detailed discussion of the latter formula can be found in Section 4.3. 
\section{B: Technical details of the test-particle simulations}
In order to calculate the turbulent magnetic field at the position of the charged particle, one can use the
Fourier representation
\be
\delta \vec{B}(x,y,z) = \int d^3 k \; \delta \vec{B} (\vec{k}) e^{i \vec{k} \cdot \vec{x}}.
\ee 
In numerical treatments of test-particle transport, the three-dimensional wavenumber integral has to be replaced
by sums. In turbulence models with reduced dimensionality such as slab or two-dimensional models, and for
isotropic turbulence, this integral can be replaced by a single sum. For the NRMHD model considered in the current
paper, however, we have to use two sums. Therefore, the turbulent magnetic field at the particle position is
given by
\be
\delta \vec{B}(x,y,z) = Re \sum_{m=1}^{N_m} \sum_{n=1}^{N_n} Amp(k_n,k_m) \; \hat{\xi}_n \exp{[i (k_n y_{n}^{\prime} + k_m z_m + \beta_n )]}.
\label{turbulence}
\ee
Here we have used the polarization vector
\bdm
\hat{\xi}_n =
\left(
\begin{array}{c}
-sin\phi_n \\
cos\phi_n \\
0
\end{array}
\right)
\label{prob}
\edm
where we have ensured that $\delta B_z = 0$. The coordinates $x_{n}^{\prime}$ and $y_{n}^{\prime}$ are obtained from a two-dimensional rotational
matrix whose azimuthal angles, $\phi_n$, are randomly generated for each summand $n$ due to symmetry reasons
\bdm
\left(
\begin{array}{c}
x_n^{'} \\
y_n^{'}
\end{array}
\right)
=
\left(
\begin{array}{cc}
-sin\phi_n & cos\phi_n \\
cos\phi_n & sin\phi_n
\end{array}
\right)
\left(
\begin{array}{c}
x \\
y
\end{array}
\right)
\edm
In Eq. (\ref{turbulence}), $Amp(k_n,k_m)=Amp(k_{\perp},k_{\parallel})$ represents the wave amplitude associated with mode $n$ and $m$.
Moreover, $k_n$ and $k_m$ stands for the wavenumbers in perpendicular and parallel directions, respectively. $\beta_n$ is just a
random plane wave phase. Basically, the NRMHD model is a broadened two-dimensional model, where the parallel component is added to the
pure perpendicular component. Therefore, the model explained above is the same as used in Hussein \& Shalchi (2014) by setting $\theta_n=\alpha_n=\pi/2$
and adding the parallel contribution separately.

The wave amplitude $Amp(k_n,k_m)$ introduced above reads
\be
Amp^2(k_n,k_m)=\frac{G(k_n)\Delta k_m \Delta k_n} {\sum_{\mu=1}^{N_m} \sum_{\nu=1}^{N_n} G(k_{\nu})\Delta k_{\mu} \Delta k_{\nu}}
\label{amp}
\ee
and the spectrum $G(k_n)$ is defined as
\be
G(k_n)=\frac{(k_nl_{\perp})^q}{[1+(k_nl_{\perp})^2]^{(s+q)/2}}.
\label{spectrum}
\ee
As in analytical treatments we have used the energy range spectral index $q$ and inertial spectral index $s$, respectively. For these two
parameters we use $q=3$ and $s=5/3$ as explained in the main part of the paper. $\Delta k_m$ and $\Delta k_n$ are the spacings between wavenumbers,
where a logarithmic spacing in $k_m$ and $k_n$ is implemented so that $\Delta k_m/k_m$ and $\Delta k_n/k_n$ are constant via the relation
\be
\frac{\Delta k_n}{k_n}= \exp\left[{\frac{ln(k_{n,max}/k_{n,min})}{N_n-1}}\right] \quad \textnormal{(same in m)}.
\label{k_step}
\ee
We should note here that $k_{m,max}=\tilde{K}$.

The trajectories of $1000$ particles where traced to yield the corresponding diffusion coefficients for each simulation run.
For the number of modes summed over in parallel and perpendicular directions, the parallel wavenumbers need to be distributed
fine enough so that the {\it resonance condition} $\mu R_L k_{\parallel} \approx 1$ is satisfied. Here we have used the unperturbed
Larmour radius $R_L$. The way how we constructed the creation of the NRMHD model in our simulations is so that we started with
a two-dimensional turbulence geometry first which only contains perpendicular wavenumbers extending theoretically till infinity.
Then we broadened this model by a parallel portion which have a cut off value at $\tilde{K}$. Taking all of that into account and
to keep computational time relatively reasonable, we have used $N_n=256$ and $N_m=32$ for our numerical calculations. It is worth
noting that we have performed test runs with $N_m$ up to $128$ and no significant differences were noticed. The size of the box was
restricted by the so-called {\it scaling condition} that ensures no particle travels beyond the maximum size of the system,
$L_{max}=k_{min}^{-1}$. This is ensured via the relation $\Omega t_{max} k_{min} R_L<1$, which corresponds to $vt_{max} < L_{max}$.
In both parallel and perpendicular direction, $k_{min}=10^{-5}$, corresponding to a relatively huge box where particles are trapped in.
Therefore we have ensured that finite box-size effects don't occur. This correspond to a spectrum without cut-off in analytical treatments
of the transport.

{}


\begin{thebibliography}{}

\bibitem[Alania et al.(2013)]{alan13}
Alania, M. V., Wawrzynczak, A., Sdobnov, V. E., \& Kravtsova, M. V. 2013, Solar Physics, 286, 561

\bibitem[Berkhuijsen et al.(2013)]{berkhuijsen2013}
Berkhuijsen, E. M., Beck, R., \& Tabatabaei, F. S. 2013, Monthly Notices of the Royal Astronomical Society, 435, 1598

\bibitem[Bieber et al.(2004)]{bieber04}
Bieber, J. W., Matthaeus, W. H., Shalchi, A., \& Qin, G. 2004, Geophys. Res. Lett., 31, 10

\bibitem[Buffie et al.(2013)]{buffie13}
Buffie, K., Heesen, V., \& Shalchi, A. 2013, The Astrophysical Journal, 764, 37

\bibitem[Engelbrecht \& Burger(2013)]{engel2013}
Engelbrecht, N. E. \& Burger, R. A. 2013, The Astrophysical Journal, 779, 158

\bibitem[Ferrand et al.(2014)]{ferrand14}
Ferrand, G., Danos, R. J., Shalchi, A., Safi-Harb, S., Edmon, P., \& Mendygral, P. 2014,
Cosmic ray acceleration at perpendicular shocks in supernova remnants, accepted for publication in ApJ,
eprint arXiv:1407.6728

\bibitem[Fyfe \& Montgomery(1976)]{fyfe76}
Fyfe, D. \& Montgomery, D. 1976, J. Plasma Phys., 16, 181

\bibitem[Fyfe et al.(1977)]{fyfe77}
Fyfe, D., Joyce, G., \& Montgomery, D. 1977, J. Plasma Phys., 17, 317

\bibitem[Ghilea et al.(2011)]{ghilea11}
Ghilea, M. C., Ruffolo, D., Chuychai, P., Sonsrettee, W., Seripienlert, A., \& Matthaeus, W. H. 2011, The Astrophysical Journal, 741, 16

\bibitem[Gradshteyn \& Ryzhik(2000)]{gra00}
Gradshteyn, I. S. \& Ryzhik, I. M., 2000, {\it Table of integrals, series, and products}, Academic Press, New York

\bibitem[Higdon(1984)]{higdon84}
Higdon, J. C. 1984, ApJ, 285, 109

\bibitem[Hussein \& Shalchi(2014)]{huss2014}
Hussein, M. \& Shalchi, A. 2014, ApJ, 785, 31

\bibitem[Jokipii(1966)]{jok66}
Jokipii, J. R. 1966, ApJ, 146, 480

\bibitem[Jokipii et al.(1993)]{jok93}
Jokipii, J. R., K\'ota, J., \& Giacalone, J. 1993, Geophysical Research Letters, 20, 1759

\bibitem[Jones et al.(1998)]{jon98}
Jones, F. C., Jokipii, J. R., \& Baring, M. G. 1998, The Astrophysical Journal, 509, 238

\bibitem[Li et al.(2012)]{Li2012}
Li, G., Shalchi, A., Ao, X., Zank, G., \& Verkhoglyadova, O. P. 2012, Advances in Space Research, 49, 1067

\bibitem[Lynn et al.(2014)]{lynn14}
Lynn, J. W., Quataert, E., Chandran, B. D. G., \& Parrish, I. J. 2014, eprint arXiv:1403.3123

\bibitem[Manuel et al.(2014)]{manuel14}
Manuel, R., Ferreira, S. E. S., \& Potgieter, M. S. 2014, Solar Physics, 289, 2207

\bibitem[Matthaeus et al.(1995)]{matt95}
Matthaeus, W. H., Gray, P. C., Pontius, D. H., Jr., \& Bieber, J. W. 1995, PhRvL, 75, 2136

\bibitem[Matthaeus et al.(2003)]{matt03}
Matthaeus, W. H., Qin, G., Bieber, J. W., \& Zank, G. P. 2003, ApJ, 590, L53

\bibitem[Matthaeus et al.(2007)]{matt2007}
Matthaeus, W. H., Bieber, J. W., Ruffolo, D., Chuychai, P., \& Minnie, J. 2007, ApJ, 667, 956

\bibitem[Minnie et al.(2009)]{minnie09}
Minnie, J., Matthaeus, W. H., Bieber, J. W., Ruffolo, D., \& Burger, R. A. 2009, Journal of Geophysical Research: Space Physics, 114, A01102

\bibitem[Montgomery(1982)]{montgomery82}
Montgomery, D. 1982, Phys. Scr., T2, 83

\bibitem[Potgieter et al.(2014)]{potgieter2014}
Potgieter, M. S., Vos, E. E., Boezio, M., De Simone, N., Di Felice, V., \& Formato, V. 2014, Solar Physics, 289, 391

\bibitem[Qin et al.(2002)]{qin2002}
Qin, G., Matthaeus, W. H., \& Bieber, J. W. 2002, Geophysical Research Letters, 29, 1048

\bibitem[Qin(2007)]{qin07}
Qin, G. 2007, The Astrophysical Journal, 656, 217

\bibitem[Ruffolo et al.(2012)]{ruffolo12}
Ruffolo, D., Pianpanit, T., Matthaeus, W. H., \& Chuychai, P. 2012, The Astrophysical Journal Letters, 747, L34

\bibitem[Ruffolo \& Matthaeus(2013)]{ruffmatt2013}
Ruffolo, D. \& Matthaeus, W. H. 2013, Physics of Plasmas, 20, 012308

\bibitem[Shalchi(2006)]{shal06}
Shalchi, A. 2006, Astronomy and Astrophysics, 453, L43

\bibitem[Shalchi \& Kourakis(2007)]{shalko07}	
Shalchi, A. \& Kourakis, I. 2007, Physics of Plasmas, 14, 112901

\bibitem[Shalchi(2009)]{shal09book}
Shalchi, A. 2009, Nonlinear Cosmic Ray Diffusion Theories, Astrophysics and Space Science Library, Vol. 362, Berlin: Springer

\bibitem[Shalchi \& Weinhorst(2009)]{Shalwei2009}
Shalchi, A. \& Weinhorst, B. 2009, Advances in Space Research, 43, 1429

\bibitem[Shalchi(2010)]{shal2010}
Shalchi, A. 2010, The Astrophysical Journal Letters, 720, L127

\bibitem[Shalchi, Li, \& Zank(2010)]{shaletal10}	
Shalchi, A., Li, G., \& Zank, G. P. 2010, Astrophysics and Space Science, 325, 99

\bibitem[Shalchi(2013a)]{shal13a}	
Shalchi, A. 2013a, Astrophysics and Space Science, 344, 187

\bibitem[Shalchi(2013b)]{shal13b}
Shalchi, A. 2013b, ApJ, 774, 7

\bibitem[Shalchi(2014)]{shal2014}
Shalchi, A. 2014, Advances in Space Research, 53, 1024

\bibitem[Snodin et al.(2013)]{snod13}
Snodin, A. P., Ruffolo, D., Oughton, S., Servidio, S., \& Matthaeus, W. H. 2013, The Astrophysical Journal, 779, 56

\bibitem[Strauss(1976)]{strauss76}
Strauss, H. R. 1976, Phys. Fluids, 19, 134

\bibitem[Tautz \& Shalchi(2011)]{tautzshal2011}
Tautz, R. C. \& Shalchi, A. 2011, ApJ, 735, 92

\bibitem[Wang et al.(2012)]{wang12}
Wang, Y., Qin, G., \& Zhang, M. 2012, The Astrophysical Journal, 752, 37

\end{thebibliography}
\end{document}